\documentclass[conference,a4paper]{IEEEtran}
\IEEEoverridecommandlockouts

\usepackage{cite}
\usepackage{dblfloatfix}
\usepackage{subcaption}
\usepackage{overpic}
\usepackage{amsmath,amssymb,amsfonts}
\usepackage{graphicx}
\usepackage{textcomp}
\usepackage{xcolor}
\usepackage{float}
\usepackage{amsthm}
\usepackage{graphicx}
\usepackage{epstopdf}
\usepackage{amsmath,bm}
\usepackage{amsfonts}
\usepackage{amssymb}
\usepackage{color}
\usepackage{multirow}
\usepackage{multicol}
\usepackage{soul,xcolor}
\usepackage{algorithm}
\usepackage{algpseudocode}

\usepackage{comment}

\theoremstyle{plain}
\newtheorem{theorem}{Theorem}

\newtheorem{lemma}{Lemma}

\newtheorem{corollary}{Corollary}

\newcommand{\vect}[1]{\mathbf{#1}}

\def\diag{\mathrm{diag}}

\def\tr{\mathrm{tr}}

\def\Htran{\mbox{\tiny $\mathrm{H}$}}
\def\Ttran{\mbox{\tiny $\mathrm{T}$}}
\def\CN{\mathcal{N}_{\mathbb{C}}} 

\def\tr{\mathrm{tr}}

\newcommand{\maximize}[1]{{\underset{{#1}}{\mathrm{maximize}}}}

\def\H{{\bf H}}
\def\F{{\bf F}}
\def\U{{\bf U}}
\def\V{{\bf V}}

\def\D{{\bf D}}
\def\G{{\bf G}}
\def\x{{\bf x}}
\def\Q{{\bf Q}}
\def\I{{\bf I}}
\def\Nt{N_{\textrm{t}}}
\def\Nr{N_{\textrm{r}}}
\def\N0{N_0}
\def\He2e{\H_{\textrm{e2e}}}
\def\Thetab{\bm{\Theta}}

\IEEEoverridecommandlockouts

\begin{document}

\title{Capacity Maximization for MIMO Channels Assisted by Beyond-Diagonal RIS}

\author{\IEEEauthorblockN{
Emil Bj{\"o}rnson\IEEEauthorrefmark{1},   
\"Ozlem Tu\u{g}fe Demir\IEEEauthorrefmark{2}   
}                                     
\IEEEauthorblockA{\IEEEauthorrefmark{1}
Department of Computer Science, KTH Royal Institute of Technology, Stockholm, Sweden, emilbjo@kth.se}
\IEEEauthorblockA{\IEEEauthorrefmark{2}
Department of Electrical-Electronics Engineering, TOBB University of Economics and Technology, Ankara, T\"urkiye}%
\thanks{E.~Bj\"ornson was supported by the  FFL18-0277 grant from the Swedish Foundation for Strategic Research. \"O. T. Demir was supported by 2232-B International Fellowship for Early Stage Researchers Programme funded by the Scientific and Technological Research Council of T\"urkiye.}
}

\maketitle
\begin{abstract}
Reconfigurable intelligent surfaces (RISs) can improve the capacity of wireless communication links by passively beamforming the impinging signals in desired directions. This feature has been demonstrated both analytically and experimentally for conventional RISs, consisting of independently reflecting elements. 
To further enhance reconfigurability, a new architecture called beyond-diagonal RIS (BD-RIS) has been proposed. It allows for controllable signal flows between RIS elements, resulting in a non-diagonal reflection matrix, unlike the conventional RIS architecture. 
Previous studies on BD-RIS-assisted communications have predominantly considered single-antenna transmitters/receivers.
One recent work provides an iterative capacity-improving algorithm for multiple-input multiple-output (MIMO) setups but without providing geometrical insights. 

In this paper, we derive the first closed-form capacity-maximizing BD-RIS reflection matrix for a MIMO channel.
We describe how this solution pairs together propagation paths, how it behaves when the signal-to-noise ratio is high, and what capacity is achievable with ideal semi-unitary channel matrices.
The analytical results are corroborated numerically.
\end{abstract}

\section{Introduction}
Reconfigurable intelligent surfaces (RISs) are arrays composed of numerous passive scattering elements that can be real-time configured to shape the wireless propagation environment to enhance communication performance \cite{li2023reconfigurable}. 
The potential role of these surfaces in next-generation wireless networks has been studied intensively for the last six years, from hardware
\cite{Tsilipakos2020a}, communications \cite{di2020smart}, and signal processing \cite{9721205} perspectives.

In a conventional RIS \cite{li2023reconfigurable}, each element is independently connected to ground and has an adjustable impedance (e.g., using a pin or varactor diode). 
This enables each RIS element to independently control how it phase-shifts an incident waveform before reflecting it. Any such RIS configuration can be represented by a diagonal reflection matrix with unit-modulus entries. Such RISs have been demonstrated experimentally~\cite{Araghi2022a}.

The RIS technology has recently been generalized into the \emph{beyond-diagonal RIS (BD-RIS)} concept, where the elements are interconnected through a passive but adjustable circuit network \cite{li2023reconfigurable,shen2021modeling}. 
The BD-RIS terminology indicates that the reflection matrix can be any unitary matrix, whereof some require non-reciprocal circuit networks (e.g., circulators). This implies that the signal impinging on an element can be divided, phase-shifted, and emitted from a multitude of elements.

A BD-RIS requires more complex circuitry than a conventional RIS, but for a given array size, it can offer higher communication performance.
The early works \cite{shen2021modeling,li2022beyond,nerini2023closed,zhou2023optimizing} studied BD-RIS-assisted narrowband single-input single-output (SISO) systems and demonstrated gains in signal-to-noise ratio (SNR) up to 62\%. The gains are largest when the elements experience varying channel conditions. The wideband SISO scenario was analyzed in \cite{10623689}, where an algorithm was proposed to maximize capacity. Recently, \cite{Santamaria2024a} presented an algorithm to maximize the capacity of narrowband multiple-input multiple-output (MIMO) channels assisted by BD-RIS. Although this algorithm performs well, the paper lacks an analytical interpretation of the optimal BD-RIS reflection matrix structure.

In this paper, we address this research gap by deriving a closed-form expression for the capacity-maximizing BD-RIS reflection matrix in a narrowband MIMO setup.
Our main theorem is supported by examples and corollaries that explore the geometrical structure, for example, how the propagation paths are paired up, what happens at high SNRs, and the ultimate capacity achieved with semi-unitary channel matrices. 
We provide numerical examples that corroborate the analytical results and benchmark the suboptimal algorithm from \cite{Santamaria2024a}. We provide appendices with detailed analytical proofs.

\vspace{-2mm}
\section{System Model and Problem Formulation}

We consider a MIMO communication system between a transmitter with $\Nt$ antennas and a receiver with $\Nr$ antennas with assistance from a BD-RIS with $M$ elements.
The goal is to derive the capacity-maximizing BD-RIS reflection matrix analytically and explain its geometrical structure. To this end, we assume perfect channel state information (CSI) and that there are no paths between the transmitter and receiver except those via the BD-RIS.
The end-to-end $\Nr \times \Nt$ MIMO channel matrix can then be expressed as
\vspace{-2mm}
\begin{equation}
\H = \F \Thetab \G^{\Htran} ,
\label{eq:MIMOchannel}
\vspace{-2mm}
\end{equation}
where $\G \in \mathbb{C}^{\Nt \times M}$ is the channel matrix from the transmitter to the BD-RIS, $\F \in \mathbb{C}^{\Nr \times M}$ is the channel matrix from the BD-RIS to the receiver, and $\Thetab \in \mathbb{C}^{M \times M}$ is the reflection matrix.
The most general form of fully connected reflective BD-RIS is considered for which the feasible set for $\Thetab$ is\footnote{Many previous works assume a reciprocal circuit network in the BD-RIS, which implies $ \Thetab = \Thetab^{\Ttran}$ \cite{shen2021modeling}. However, since non-reciprocal circuit networks can be built (e.g., using circulators), we omit this constraint to establish the ultimate capacity limit. The same was recently done in \cite{Li2024a}.}
\begin{equation}
\mathcal{U} =  \left\{ \Thetab  :  \Thetab^{\Htran} \Thetab = \I_M \right\}.
\end{equation}
This enables a signal impinging on one element to be partially reflected from another element, in a lossless and non-reciprocal manner.
For any selection of $\Thetab$, we obtain a conventional MIMO channel, and its capacity is achieved by sending a complex Gaussian signal  $\x \sim \CN({\bf 0}, \Q)$, where the positive semi-definite covariance matrix $\Q \succeq {\bf 0}$ has eigenvectors equalling the right singular vectors of $\H$ and eigenvalues computed based on the waterfilling algorithm \cite{bjornson2024introduction}.
We want to jointly optimize $\Q$ with the reflection matrix to achieve the largest possible capacity, which can be formulated as follows:
\begin{align}
\maximize{\Thetab, \Q}\,\,& \log_2 \det \left( \I_{N_R} + \frac{1}{\N0}\H \Q \H^{\Htran} \right) \label{eq:Capacity}  \\
\mathrm{subject\,to}\,\,& \,\tr(\Q) \leq q_{\max}, \quad  \Q \succeq {\bf 0}, \\
\,\,& \,\Thetab \in \mathcal{U}  ,
\end{align}
where $q_{\max}$ is the total transmit power and $\N0$ is the power of the independent additive complex Gaussian receiver noise.

\section{Capacity Maximizing BD-RIS Configuration}

In this section, we will solve the maximization problem in \eqref{eq:Capacity} and analyze the solution.
We can express the singular value decompositions (SVDs) of the channel matrices as
\vspace{-2mm}
\begin{align}
\F &= \U_{F}   \D_{F} \V_{F}^{\Htran} \\
\G &= \U_{G}  \D_{G}  \V_{G}^{\Htran}, 
\end{align}
where $\D_{F} \in \mathbb{C}^{\Nr \times M}, \D_{G} \in \mathbb{C}^{\Nt \times M}$ are rectangular diagonal matrices containing the singular values in decreasing order, 
the columns of $\U_{F} \in \mathbb{C}^{\Nr \times \Nr}, \U_{G} \in \mathbb{C}^{\Nt \times \Nt}$ contain the corresponding left singular vectors, and the columns of 
$\V_{F}  \in \mathbb{C}^{M \times M}, \V_{G} \in \mathbb{C}^{M \times M}$ contain the corresponding right singular vectors.
Furthermore, we let $K = \min(\Nt,\Nr,M)$ and let $\sigma_i(\cdot)$ denote the $i$th largest singular value of a matrix.
Using this notation, we can state the following result, which is aligned with the results obtained in \cite[Eqn. (16)]{do2022line} and \cite[Prop.~1]{bartoli2023spatial} but proved differently.
\vspace{-2mm}

\begin{theorem} \label{th:capacity}
The optimization problem  \eqref{eq:Capacity} is solved by\footnote{A (block-)diagonal matrix with $a_1,\ldots,a_N$ is denoted by $\diag(a_1,\ldots,a_N)$.}
\begin{align} \label{eq:optimalTheta}
\Thetab &= \V_{F} \V_{G}^{\Htran}  \\
\Q &= \U_{G} \diag(q_1,\ldots,q_{\Nt}) \U_{G}^{\Htran},  \label{eq:optimalQ}
\end{align}
and the maximum capacity value is
\begin{equation}
C = \sum_{i=1}^{K} \log_2 \left( 1+\frac{q_i \sigma_i^2 (\F) \sigma_i^2 (\G^{\Htran}) }{\N0} \right), \label{eq:capacity}
\end{equation}
where the transmit powers $q_1,\ldots,q_{K}$ are computed using the waterfilling algorithm: $q_i = \mu - \N0  / ( \sigma_i^2 (\F) \sigma_i^2 (\G^{\Htran}) )$, where $\mu$ is selected so that $\sum_{i=1}^{K} q_i = q_{\max}$.
\end{theorem}
\begin{IEEEproof}
The proof is given in Appendix~\ref{app:proofCapacity}.
\end{IEEEproof}

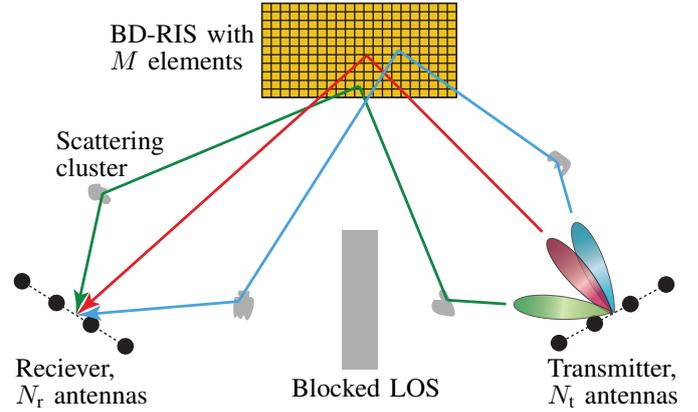
\begin{figure}[t!]
	\centering 
	\begin{overpic}[width=\columnwidth,tics=10]{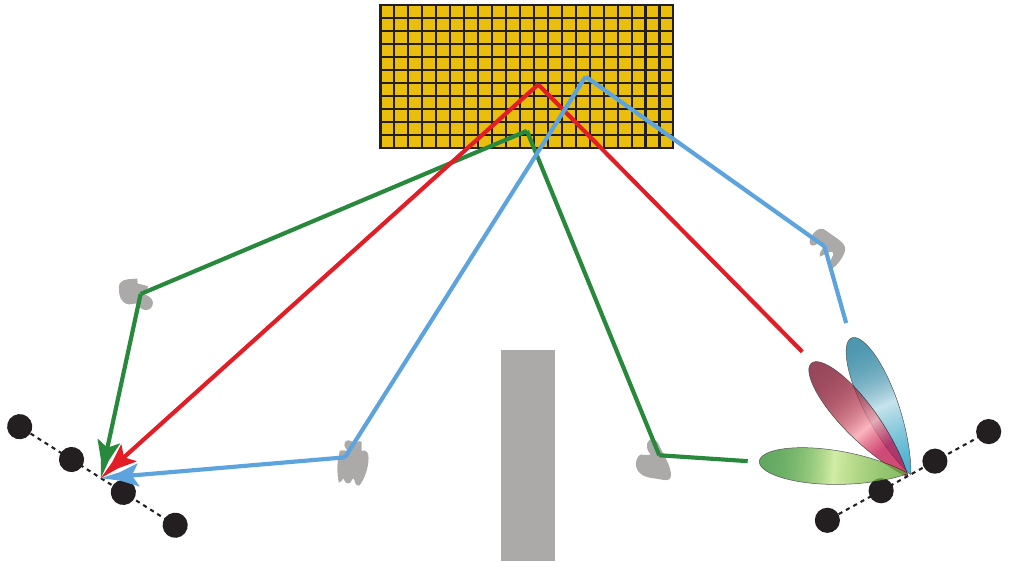}
	 \put (15,49) {BD-RIS with}
	 \put (15,45) {$M$ elements}
	 \put (42,-4) {Blocked LOS}
	 \put (7,33) {Scattering}
	 \put (7,29) {cluster}
  \put (80,-1) {Transmitter,}
    \put (80,-5) {$\Nt$ antennas}
    \put (1,-1) {Reciever,}
        \put (1,-5) {$\Nr$ antennas}
\end{overpic} \vspace{0.1mm}
	\caption{Theorem~\ref{th:capacity} proves that the optimal BD-RIS configuration will reflect one transmitter-RIS path along one RIS-receiver. The pairing is made in the order of strength.}
	\label{fig:MIMO_sketch}
    \vspace{-6mm}
\end{figure}

The key insight from this theorem is that the optimal BD-RIS configuration combines the $i$th strongest singular value directions of $\F$ and $\G$, for $i=1,\ldots,K$. A geometrical interpretation is provided in Fig.~\ref{fig:MIMO_sketch}, where there are three paths between the transmitter (or receiver) and the BD-RIS, each corresponding to one singular value of $\G$ (or $\F$). The optimal reflection matrix will connect the strongest paths (i.e., the line-of-sight (LOS) paths) and then continue pairing up paths in a decaying order of strength. The geometrical structure is independent of $q_{\max}$, but the waterfilling determines how many of the paired-up paths will be allocated non-zero signal power.

The capacity-maximizing BD-RIS configuration in Theorem~\ref{th:capacity} is not unique. If the waterfilling only assigns non-zero power to $\tilde{K}$ singular values, then the essential thing is that the BD-RIS pairs up the $\tilde{K}$ strongest paths in $\vect{F}$ and $\vect{G}$ in decaying order. The channel dimensions not used for signaling can be treated arbitrarily by using $\Thetab = \V_{F} \diag(\vect{I}_{\tilde{K}},\tilde{\vect{U}}) \V_{G}^{\Htran}$, where $\tilde{\vect{U}}$ can be any $(M-\tilde{K}) \times (M-\tilde{K})$ unitary matrix.

When the SNR is high, the set of capacity-maximizing  BD-RIS configurations increases even further, which can be formalized as follows.

\vspace{-2mm}
\begin{corollary} \label{corollary} 
Suppose $\vect{F}$ and $\vect{G}$ have $K$ non-zero singular values.
At high SNR, where the waterfilling power allocation becomes $q_i= q_{\rm max}/K$ for $i=1,\ldots,K$, the maximum capacity is achieved by all BD-RIS configurations of the kind
\begin{align} \label{eq:optimalTheta2}
\Thetab &= \V_{F} \vect{P}\V_{G}^{\Htran},
\end{align}
where $\vect{P}$ is a block-diagonal matrix with the first block being any $K \times K$ permutation matrix and the second block being any $(M-K) \times (M-K)$ unitary matrix.
\end{corollary}
\begin{IEEEproof}
The proof is given in Appendix~\ref{app:proofHighSNR}.
\end{IEEEproof}

This corollary demonstrates that the optimal configuration at high SNR maps the $K$ largest singular values of $\vect{F}$ to the $K$ largest singular values of
$\vect{G}$, but the ordering can be arbitrary.

In the next section, we will analyze the behavior of channel matrices with orthogonal columns/rows and equal singular values, and examine when a conventional RIS can provide the same capacity as a BD-RIS when assisting a MIMO system. 

\subsection{On the capacity of channels with equal singular values}

An ideal MIMO channel has many equally strong paths. 
Hence, we consider the case when $\vect{F}$ and $\vect{G}$ are semi-unitary matrices that each have equal singular values, $\sigma_i(\vect{F})=\sigma_{\rm F}$ and $\sigma_i(\vect{G})=\sigma_{\rm G}$, for $i=1,\ldots,K$. Then, \eqref{eq:capacity} becomes
\begin{equation}
C = \sum_{i=1}^{K} \log_2 \left( 1+\frac{q_i \sigma_{\rm F}^2 \sigma_{\rm G}^2}{\N0} \right) = K \log_2 \left( 1+\frac{q_{\rm max} \sigma_{\rm F}^2 \sigma_{\rm G}^2}{K\N0} \right), \label{eq:capacity2}
\end{equation}
where we utilized that the equal allocation $q_i=q_{\rm max}/K$ is optimal.
We can observe from \eqref{eq:capacity2} that the spatial multiplexing gain is $K = \min(M, \Nt, \Nr)$. When $M < \min(\Nt, \Nr)$, the spatial multiplexing gain is constrained by the number of elements in the BD-RIS. We then have the following result.

\begin{corollary} \label{th:orthogonal}
Suppose $\vect{F}$ and $\vect{G}$ are semi-unitary matrices, each having $K$ equally large singular values. If  $M \leq \max(\Nr, \Nt)$, then the maximum capacity is achieved using any BD-RIS configuration.
\end{corollary}
\begin{IEEEproof}
The proof is given in Appendix~\ref{app:proofOrthogonal}.
\end{IEEEproof}

This corollary identifies a special case when any unitary $\vect{\Theta}$ works equally well. An important consequence of Corollary~\ref{th:orthogonal} is that a conventional RIS with an arbitrary phase-shift configuration is sufficient to maximize the capacity. By contrast, we expect a BD-RIS to outperform an RIS in any other scenario (e.g., with $M > \max(\Nr, \Nt)$ or with varying singular values).

\section{Numerical Results}
In this section, we will evaluate the capacity of a BD-RIS-assisted MIMO channel. We will compare the closed-form optimal configuration from Theorem \ref{th:capacity} with the iterative algorithm that was proposed in \cite{Santamaria2024a}. The latter one finds a good BD-RIS reflection matrix under the additional reciprocity constraint $\vect{\Theta} = \vect{\Theta}^{\Ttran}$. Both solutions are theoretically implementable, but our solution provides a higher capacity at the cost of requiring non-reciprocal components (e.g., circulators).

We model $\vect{F}$ and $\vect{G}$ with Rician fading using a $k$-factor of $3$\,dB.  Both the transmitter and receiver are equipped with uniform linear arrays (ULAs) with $\lambda/2$ spacing, where $\lambda$ is the wavelength. The BD-RIS is deployed as a uniform planar array (UPA) with $\lambda/4$ spacing. The LOS paths are generated with randomly selected azimuth and elevation angles uniformly distributed in $[-\pi/2,\pi/2]$.
Similarly, the non-LOS paths of each channel matrix are generated using $20$ clusters with random azimuth and elevation angles and fading coefficients, assuming equal average power distribution among the clusters. We will specify the ``per-element'' SNR that is obtained in a reference setup with $\Nt=\Nr=M$. The SNR is $-10$\,dB unless otherwise stated. The number of receive and transmit antennas is $\Nr=\Nt=16$. Each point on the simulation curves is computed as the average over many independent channel realizations with random angles and clusters.

\begin{figure}[t] 
    \centering
        \includegraphics[width=0.5\textwidth, trim=0.6cm 0.2cm 0.8cm 0cm, clip]{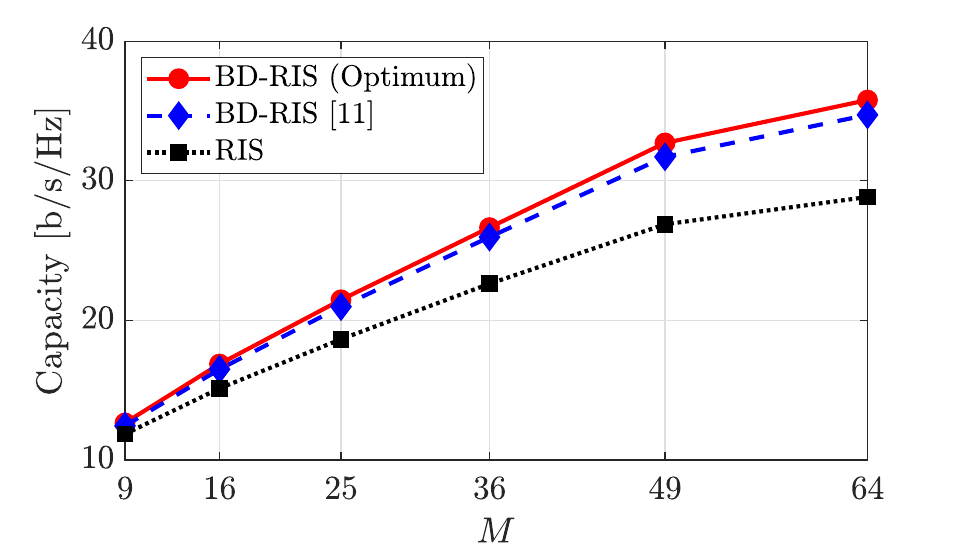} 
        \caption{The capacity achieved with different RIS architectures and configurations versus the number of elements ($M$).}
        \label{fig:1}
        \vspace{-4mm}
 \end{figure}

In Fig.~\ref{fig:1}, we present the capacity achieved by various BD-RIS and RIS configurations. The ``BD-RIS (Optimum)'' curve shows the performance obtained by the closed-form solution in Theorem \ref{th:capacity}. The capacity obtained using the algorithm from \cite{Santamaria2024a}, which computes a reciprocal BD-RIS configuration, is also shown.\footnote{We have empirically observed that initializing the algorithm in \cite{Santamaria2024a} with $\vect{F}^{\Htran}\vect{H}^{\star}\vect{G}$, where $\vect{H}^{\star}$ is the end-to-end channel matrix obtained with the conventional RIS configuration in \cite[Alg.~9.2]{bjornson2024introduction}, results in faster convergence and higher capacity (especially at high SNR) than the original initialization method described in \cite{Santamaria2024a}. Hence, we have used this initialization in this paper.}
The performance of a conventional RIS is used as a baseline, and its configuration is optimized using \cite[Alg.~9.2]{bjornson2024introduction}. 
The results demonstrate that both BD-RIS architectures outperform the conventional RIS, particularly when the number of elements ($M$) is large.
The maximum capacity is achieved by the BD-RIS configuration proposed in this paper, which is expected since it is provably optimal.
The performance gap between Theorem \ref{th:capacity} and the algorithm in \cite{Santamaria2024a}  is likely due to the latter only considering reciprocal BD-RIS scattering networks. 
Moreover, that algorithm might not converge to the optimal reciprocal solution, but this is a minor issue since the performance gap is small---though it increases with $M$.
In summary, the proposed theorem quantifies the maximum achievable capacity, and similar values can be achieved under additional constraints such as reciprocity.

    \begin{figure}[t] 
    \centering
        \includegraphics[width=0.5\textwidth, trim=0.2cm 0.2cm 0.8cm 0cm, clip]{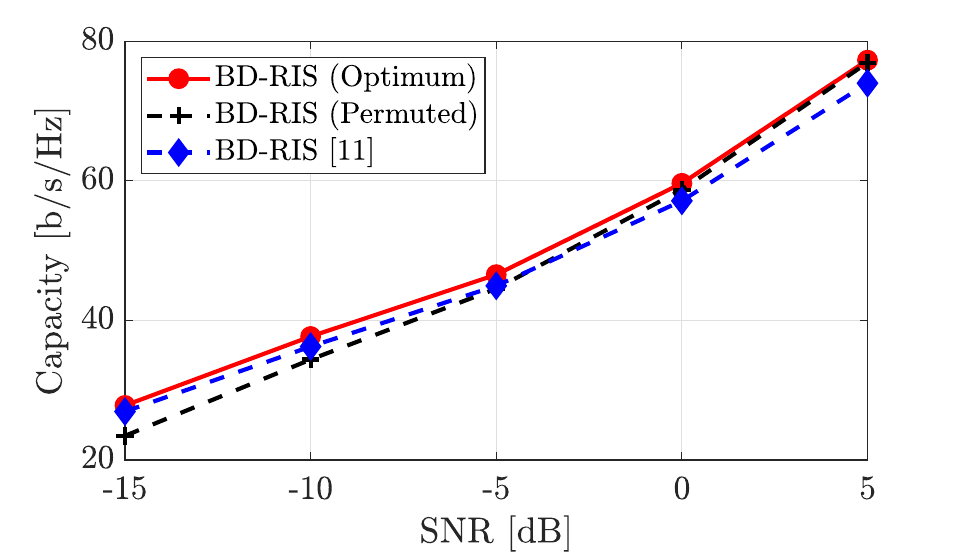} 
        \caption{The capacity achieved versus the SNR for $M=64$.}
        \label{fig:2}
        \vspace{-4mm}
 \end{figure}

To validate Corollary~\ref{corollary}, Fig.~\ref{fig:2} shows the capacity as a function of the reference SNR when $M=64$. In addition to the optimal configuration and the algorithm from \cite{Santamaria2024a}, we also include the average performance obtained by a BD-RIS configuration generated using a random permutation matrix in $\vect{P}$ (defined as in the corollary). In this case, $K$ is chosen as the minimum number such that the sum of the $K$ largest singular values is at least 95\% of the sum of all singular values, and this holds for both $\vect{F}$ and $\vect{G}$. This scheme is labeled ``Permuted.'' At low SNR, there is a noticeable gap between the optimal and permuted solutions. 
However, as the SNR increases, the performance gap vanishes, demonstrating the correctness of Corollary~\ref{corollary}, which states that we can combine the $K$ strongest directions in $\vect{F}$ and $\vect{G}$ arbitrarily when the SNR is large.

     \begin{figure}[t] 
    \centering
        \includegraphics[width=0.5\textwidth, trim=0.2cm 0.2cm 0.8cm 0cm, clip]{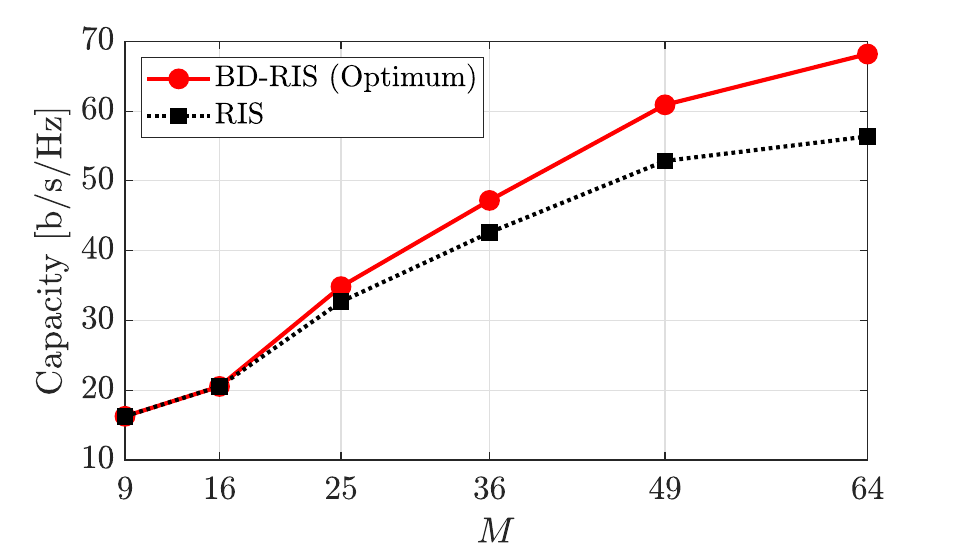} 
        \caption{The capacity achieved versus the number of elements ($M$) when having semi-unitary channels.}
        \label{fig:3}
          \vspace{-4mm}
 \end{figure}

 Finally, in Fig.~\ref{fig:3}, we consider semi-unitary channel matrices $\vect{F}$ and $\vect{G}$ by replacing the original channels' singular values with equal values. 
The figure shows that when $M \leq \Nr = \Nt = 16$, the capacities obtained with the conventional RIS and the optimal BD-RIS configuration are identical. 
 This is aligned with Corollary \ref{th:orthogonal}.
However, when $M$ exceeds $16$, a performance gap emerges due to the mismatch between the subspaces spanned by $\vect{F}$ and $\vect{G}$. This mismatch can be managed by configuring the BD-RIS to map the subspace to each other, while it can only be partially handled by selecting the diagonal phase-shift matrix in the conventional RIS.

\section{Conclusions}

The BD-RIS architecture is significantly more complex than the conventional RIS architecture; thus, it is essential to identify in which situations the performance gap is also significant.
While previous works have focused on SISO channels, this paper considers a BD-RIS-assisted MIMO channel. 
We derived the BD-RIS configuration that maximizes the capacity. Since the solution is in closed form, it reveals that the BD-RIS should pair up singular vector dimensions between the transmitter$\to$BD-RIS and BD-RIS$\to$receiver channels in the same order of strength. The waterfilling power allocation then determines how many of these paths are used for signal transmission but does not change the BD-RIS configuration.

When the SNR is high, there is more flexibility in the optimal BD-RIS configuration. The strongest singular vector dimensions from the two channel matrices should still be utilized but can be paired up arbitrarily. Additionally, we examined semi-unitary channel matrices, which are ideal for MIMO communications.
If the number of elements is not greater than the number of antennas at the transmitter and receiver, a conventional RIS achieves the same performance as the optimal BD-RIS configuration.
By contrast, the numerical results showed that the capacity gains provided by the BD-RIS architecture become larger as the number of elements ($M$) increases, far beyond the number of antennas.
The optimal non-reciprocal solution and the reciprocal solution from \cite{Santamaria2024a} provide similar capacity, but there is a gap when $M$ is large.

\appendices
\vspace{-2mm}
\section{Proof of Theorem~\ref{th:capacity}}
\label{app:proofCapacity}
\vspace{-2mm}

The proof builds on the following lemma.

\begin{lemma} \label{lemma:prodSingular}
For matrices $\vect{A}_1,\ldots,\vect{A}_p \in \mathbb{C}^{M \times M}$ it holds that
\begin{align} \nonumber
\sum_{i=1}^{m} & \log_2 \left( 1 + \sigma_i (\vect{A}_1 \cdots \vect{A}_p)  \right) \\  &\leq \sum_{i=1}^{m} \log_2 \left( 1 + \sigma_i (\vect{A}_1) \cdots \sigma_i(\vect{A}_p)  \right) \label{eq:prodSingular_ineq}
\end{align}
for $m=1,\ldots,M$, with strict equality for $m=M$.
\end{lemma}
\begin{IEEEproof}
For the given set of matrices, \cite[Th.~H.1.b]{Marshall2011a} states that 
$\prod_{i=1}^{m}  \sigma_i (\vect{A}_1 \cdots \vect{A}_p) \leq \prod_{i=1}^{m}  \sigma_i (\vect{A}_1) \cdots \sigma_i(\vect{A}_p)$ for $m=1,\ldots,M$ (with strict equality for $m=M$).
This inequality can then be turned into \eqref{eq:prodSingular_ineq} by applying \cite[Th.~3.3.10(b)]{Horn1991a} with the function $f(x) = \log_2(1+x)$.
\end{IEEEproof}

For notational convenience, we will prove Theorem~\ref{th:capacity} for the case of $M \geq \min(\Nt,\Nr)=K$.
For any given $\Thetab$ and any $\vect{Q}$ with the ordered singular values denoted as $q_1,\ldots,q_{\Nt}$, the objective function in \eqref{eq:Capacity} can be expressed as
\vspace{-2mm}
\begin{align} \nonumber
& \log_2 \det \left( \I_{N_R} + \frac{1}{\N0}\H \Q \H^{\Htran} \right) \\ \nonumber &= 
\log_2 \det \left( \I_{N_R} + \frac{1}{\N0} \F \Thetab \G^{\Htran} \Q \G \Thetab^{\Htran} \F^{\Htran} \right) \\ \nonumber &= 
\log_2 \det \bigg( \I_{M} +  \underbrace{\begin{bmatrix} \F \\ \vect{0} \end{bmatrix}}_{=\vect{A}_1}  \underbrace{\begin{bmatrix} \G \Thetab^{\Htran} \\ \vect{0} \end{bmatrix}^{\Htran}}_{=\vect{A}_2} \! \underbrace{\begin{bmatrix}  \frac{1}{\N0}  \Q & \vect{0} \\ \vect{0} & \vect{0} \end{bmatrix}}_{=\vect{A}_3} \! \underbrace{\begin{bmatrix} \G \Thetab^{\Htran} \\ \vect{0} \end{bmatrix}}_{=\vect{A}_4} \! \underbrace{\begin{bmatrix} \F \\ \vect{0} \end{bmatrix}^{\Htran}}_{=\vect{A}_5} \bigg) \\
&= \sum_{i=1}^{K}  \log_2 \left( 1 + \sigma_i (\vect{A}_1 \cdots \vect{A}_5)  \right),
\end{align}
where the matrices $\vect{A}_1,\ldots,\vect{A}_5$ are padded with zeros to become $M \times M$, but we only included $K=\min(\Nt,\Nr)$ terms in the last expression since the remaining singular values are zero.
It then follows from Lemma~\ref{lemma:prodSingular} that 
\begin{align} \nonumber
& \log_2 \det \left( \I_{N_R} + \frac{1}{\N0}\H \Q \H^{\Htran} \right) \\ \nonumber & \leq 
\sum_{i=1}^{K} \log_2 \left( 1 + \sigma_i (\vect{A}_1) \cdots \sigma_i(\vect{A}_5)  \right)  \\
&= \sum_{i=1}^{K} \log_2 \left( 1 + \frac{1}{N_0} q_i \sigma_i^2 (\vect{F})   \sigma_i^2(\vect{G}^{\Htran})  \right) \label{eq:proof-upper-bound}
\end{align}
by utilizing that $\sigma_i (\vect{Q}) = q_i$ and $  \sigma_i^2( \Thetab \vect{G}^{\Htran}) =   \sigma_i^2(\vect{G}^{\Htran})$ since $\Thetab$ is a unitary matrix. The upper bound in \eqref{eq:proof-upper-bound} is achieved if we select 
$\Thetab$ and $\vect{Q}$ according to \eqref{eq:optimalTheta}--\eqref{eq:optimalQ} because this makes
\begin{align} \nonumber
&\H \Q \H^{\Htran} = \F \Thetab \G^{\Htran} \Q \G \Thetab^{\Htran} \F^{\Htran} \\
& =\U_{F} \D_{F}  \D_{G}^{\Htran} \diag(q_1,\ldots,q_{\Nt}) \D_{G}  \D_{F}^{\Htran} \U_{F}^{\Htran},
\end{align}
for which the $i$th singular value is $q_i \sigma_i^2 (\vect{F})   \sigma_i^2(\vect{G}^{\Htran})$ for $i=1,\ldots,K$.
As this result holds for any values of $q_1,\ldots,q_{\Nt}$, we can further increase the rate by using waterfilling. This concludes the proof for the case of $M \geq \min(\Nt,\Nr)$.

If $M <  \min(\Nt,\Nr)$, we can follow the same proof steps but must pad $\vect{A}_1,\ldots,\vect{A}_5$ with zeros differently since $K=M$. 

\section{Proof of Corollary~\ref{corollary}}
\label{app:proofHighSNR}
At high SNR (i.e., $q_{\rm max}$ is large), a tight lower bound on the capacity expression in Theorem~\ref{th:capacity} is
\begin{align} \nonumber
C &= \sum_{i=1}^{K} \log_2 \left( \frac{q_{\rm max} \sigma_i^2 (\F) \sigma_i^2 (\G^{\Htran}) }{K \N0} \right) \\
&=  \log_2 \left(  \left( \frac{q_{\rm max} }{K \N0} \right)^K  \prod_{i=1}^{K}  \sigma_i^2 (\F) \sigma_i^2 (\G^{\Htran}) \right).
\end{align}
Since this expression depends on the product of the singular values, it also equals
\begin{equation}
    C = \sum_{i=1}^{K} \log_2 \left( \frac{q_{\rm max} \sigma_i^2 (\F) \sigma_{p_i}^2 (\G^{\Htran}) }{K \N0} \right),
\end{equation}
where $p_1,p_2,\ldots,p_K$ can be any permutation of $1,2,\ldots,K$. Hence, we can also maximize the high-SNR capacity using $\Thetab = \V_{F} \vect{P}\V_{G}^{\Htran}$, where $\vect{P}$ is a block-diagonal matrix containing the permutation matrix 
representing $p_1,p_2,\ldots,p_K$ and an arbitrary unitary $(M-K) \times (M-K)$ matrix.

\section{Proof of Corollary~\ref{th:orthogonal}}
 \label{app:proofOrthogonal}

 There are three cases to be evaluated: i) $M\leq \Nr$ and $M\leq \Nt$; ii) $\Nt\leq M\leq \Nr$; and iii) $\Nr \leq M \leq \Nt$. We will analyze each case in the sequel and denote the singular values of $\vect{F}$ and $\vect{G}$ as $\sigma_i(\vect{F})=\sigma_{\rm F}$ and $\sigma_i(\vect{G})=\sigma_{\rm G}$, for $i=1,\ldots,K$.

 {\bf Case 1} ($M\leq \Nr$ and $M\leq \Nt$): We have $K=M$ so the compact SVD of $\vect{H}$ (for any unitary $\vect{\Theta}$) can be expressed as
 \begin{align}
 \vect{H} =  \underbrace{\frac{1}{\sigma_{\rm F}} \vect{F}\vect{\Theta}}_{=\vect{U}_H} \underbrace{\sigma_{\rm F} \sigma_{\rm G}\vect{I}_K}_{=\vect{D}_H} \underbrace{\frac{1}{\sigma_{\rm G}} \vect{G}^{\Htran}}_{=\vect{V}_H^{\Htran}}.
 \end{align}
The singular values in $\vect{D}_H$ are all equal and unaffected by the choice of $\vect{\Theta}$, which proves the result in this case.

 {\bf Case 2} ($\Nt\leq M\leq \Nr$):
We have $K=\Nt$, $\vect{F}^{\Htran}\vect{F}=\sigma_{\rm F}^2\vect{I}_M$ and $\vect{G}\vect{G}^{\Htran}=\sigma_{\rm G}^2\vect{I}_{\Nt}$. If we use equal power allocation with $\vect{Q}=\frac{q_{\rm max}}{K}\vect{I}_K$, the rate in \eqref{eq:Capacity} becomes
\begin{align} \nonumber
    &\log_2 \det \!\left(\vect{I}_{\Nr}+\frac{q_{\rm max}}{KN_0}\vect{H}\vect{H}^{\Htran}\right) \!= \log_2 \det \!\left(\vect{I}_{\Nt}+\frac{q_{\rm max}}{KN_0}\vect{H}^{\Htran}\vect{H}\right) \\
    &= \log_2\det\left(\vect{I}_{\Nr}+\frac{q_{\rm max}\sigma_{\rm F}^2\sigma_{\rm G}^2}{KN_0}\vect{I}_{\Nr}\right) 
\end{align}
where the first equality follows from Sylvester's determinant identity and the second equality utilizes the fact that
\begin{align}\vect{H}^{\Htran}\vect{H}=\vect{G}\vect{\Theta}^{\Htran}\vect{F}^{\Htran}\vect{F}\vect{\Theta}\vect{G}^{\Htran} = \sigma_{\rm F}^2\vect{G}\vect{G}^{\Htran} = \sigma_{\rm F}^2 \sigma_{\rm G}^2\vect{I}_{\Nt}
\end{align}
for any unitary $\vect{\Theta}$. The value coincides with \eqref{eq:capacity2}, so the maximum capacity can be achieved for any $\vect{\Theta}$ in this case.

{\bf Case 3} ($\Nr\leq M\leq \Nt$):
By using the fact that the capacity of the channel $\vect{H}$ equals the capacity of the channel $\vect{H}^{\Ttran}$, we can repeat the steps in Case 2 to prove the desired result.
 
\bibliographystyle{IEEEtran}
\bibliography{IEEEabrv,refs}

\end{document}